\documentclass[aps,prb,10pt,twocolumn,groupedaddress]{revtex4-2}

\usepackage[separate-uncertainty=true]{siunitx}
\usepackage{graphicx}
\usepackage{here}
\usepackage{amsmath}
\usepackage{nicefrac}
\usepackage{xcolor}
\usepackage{color}

\usepackage{dcolumn}
\usepackage{multirow}
\usepackage[normalem]{ulem}
\usepackage{tabto}
\usepackage{afterpage}
\DeclareSIUnit{\atpercent}{at\%}
\DeclareSIUnit{\evolt}{eV}
\DeclareSIUnit\angstrom{\text {Å}}

\begin{document}
\renewcommand{\arraystretch}{2}


\title{Annealing time, temperature, and field dependence of pinned magnetic moments in the collinear antiferromagnet PdMn}

\author{Nicolas Josten}
\email{Nicolas.Josten@uni-due.de}
\affiliation{Faculty of Physics and Center for Nanointegration (CENIDE), University Duisburg Essen, Duisburg, 47057, Germany}
\author{Olga Miroshkina}
\author{Mehmet Acet}
\author{Markus Gruner}
\author{Michael Farle}
\affiliation{Faculty of Physics and Center for Nanointegration (CENIDE), University Duisburg Essen, Duisburg, 47057, Germany}


\date{\today}

\begin{abstract}
Magnetic annealing of the collinear antiferromagnet PdMn with excess Pd produces strongly pinned magnetic moments in the annealing field direction. This behaviour can be understood with the help of the magnetic-field-biased diffusion model. Here, the magnetic field creates an energy difference between the two possible occupations of the antiferromagnetic Mn-sublattices by the Pd-excess atoms. This, mediated by diffusion, leads to an imbalance in the amount of the Pd-excess atoms in these sublattices and, subsequently, to an imbalance in the total magnetization of the sublattices. We~investigate this effect’s dependence on the annealing field, time, and temperature. The results are then compared to the results of the magnetic-field-biased diffusion model, which gives good agreement.     
\end{abstract}


\maketitle

\section{Introduction}
Mn and its alloys are known for their versatile and interesting magnetic properties. These include the complex magnetism of manganese in all allotropic forms \cite{kasper1956antiferromagnetic, yamada1970magnetism}, spin-glasses (e.g. Cu-Mn) \cite{binder1986spin}, and Heusler alloys (e.g. Cu$_2$MnAl) \cite{heusler1903magnetische}. Another class of Mn-alloys are the collinear antiferromagnets IrMn, PdMn, PtMn, NiMn, and RhMn, which have a tetragonal L1$_0$-structure and are often used as the pinning layer in exchange biased films \cite{fukamichi2006magnetic,berkowitz1999exchange}.

It was shown that annealing NiMn with up to \SI{5}{\percent} excess Ni under a magnetic field leads to the formation of strongly pinned uncompensated magnetic moments \cite{Pal}. This effect manifests itself as a shift along the magnetization axis in the magnetic field dependence of the magnetization $M(B)$. The mechanism is thought to be related to magnetic-field-biased diffusion \cite{Pal,NiMnpaper}. Excess Ni-atoms have to occupy Mn-positions. Since NiMn is a collinear antiferromagnet, excess Ni has to occupy one or the other of the two antiferromagnetic (AF) sublattices, which results in locally induced magnetic moments. During magnetic annealing under an annealing field ($B_\text{a}$) applied in a given direction, the occupation of one of these sublattices is energetically more favorable for the excess Ni-atoms. This leads to an imbalance in the distribution among the sublattices and subsequently to a net magnetization.

In certain cases, Pd can have a ferromagnetic (FM) coupling with its environment \cite{crangle1965dilute,vitos2000size}. Therefore, it can be expected that excess Pd exhibits FM ordering induced by a Mn environment in Pd-rich PdMn as excess Ni does in Ni-rich NiMn.

PdMn around the equiatomic composition transforms martensitically between a high-temperature B2 phase and a low-temperature L1$_0$ phase \cite{hansen1958constitution}. At the equiatomic composition, the transition temperature is about \SI{900}{\kelvin} \cite{grube1936magnetische,hansen1958constitution,kjekshus1967equiatomic}. The L1$_0$ phase of PdMn is a collinear antiferromagnet, and its N\'eel temperature lies around \SI{820}{\kelvin} \cite{kren1966neutron,kjekshus1967equiatomic,Pal2}. At the equiatomic composition, only the Mn atoms carry a magnetic moment, which is around 4~$\mu_{\rm B}$ found in the results of both experimental \cite{kren1966neutron,kjekshus1967equiatomic} and \textit{ab initio} theoretical studies \cite{umetsu2002electrical,wang2012structural}. The magnetic easy-axis lies within the c-plane \cite{kren1966neutron}, and the calculated magnetocrystalline anisotropy constant is $K_2$=\SI{-1.53e6}{\joule\per\cubic\meter} \cite{Umetsu-2006}.

The aim of this work is to show that magnetic annealing of PdMn with excess Pd leads to local FM coupling and pinning of the magnetization along $B_\text{a}$. For this, we present the dependencies of the amount of pinned magnetization on $B_\text{a}$, the annealing time ($t_\text{a}$), and the annealing temperature ($T_\text{a}$) and use the magnetic-field-biased diffusion model to describe the results.

\section{Methods}

\subsection{Experimental}

PdMn with excess Pd was prepared by arc melting of pure elements (99.98\%). The sample was homogenized for 5 days at \SI{1073}{\kelvin} encapsulated in a quartz tube under argon atmosphere. Afterwards, it was quenched in water at room temperature. We used energy-dispersive X-ray spectroscopy (EDX) incorporated in a scanning electron microscope and determined the composition as Pd$_{52.9}$Mn$_{47.1}$. The sample was cut into disks with ${\sim\SI{1}{\milli\meter}}$ thickness using a precision sectioning saw. X-ray diffraction (XRD) measurements were carried out using Cu K-$\alpha$ radiation. Each disk was cut into multiple cuboids ranging from \SI{19.9}{\milli\gram} to \SI{59.7}{\milli\gram} for magnetization measurements. Additional grinding removed any cutting residues.

The magnetic properties were measured in a Quantum Design MPMS XL superconducting quantum interference device (SQUID) magnetometer with a high-temperature unit. First, a \SI{5}{\tesla}-$M(B)$-curve at \SI{300}{\kelvin} was taken for an initial state sample. The field was then set to the value of $B_\text{a}$ and warmed to $T_\text{a}$ at a rate of \SI{4}{\kelvin\per\minute}. The sample remained at this temperature for a chosen time and subsequently cooled back to \SI{300}{\kelvin} at the same rate without removing the field. Afterwards, $M(B)$ was measured again at \SI{300}{\kelvin}. $M(B)$ at \SI{300}{\kelvin} of the annealed state shows a vertical shift with respect to $M(B)$ of the initial state (as-prepared). This procedure was repeated for various $B_\text{a}$, $t_\text{a}$, and $T_\text{a}$ to understand the mechanisms behind the occurrence of the vertical shift. A fresh sample was used for each $B_\text{a}$ and $T_\text{a}$. For the $t_\text{a}$-dependence of the vertical shift, the same sample was used for consecutive annealing steps.

\subsection{\textit{Ab initio} calculations}
Density functional theory~(DFT) calculations were performed using the Vienna \textit{Ab~Initio} Simulation Package~(VASP)~\cite{Kresse-1996, Kresse-1999} with the generalized gradient approximation~(GGA) in the Perdew, Burke, and Ernzerhof~(PBE)~formulation~\cite{Perdew-1996} and a scalar-relativistic approximation with collinear spin alignment of magnetic moments.
We~employed PAW potentials with the following valence states: 
$3p^6 3d^5 4s^2$ for Mn,
$4p^6 4d^{10}$ for Pd, and
$3p^6 3d^8 4s^2$ for~Ni (for the comparison with Ni-Mn, see Appendix~\ref{Appendix_DFT}).
A supercell approach was implemented to simulate the equiatomic PdMn and Pd-excess PdMn alloys. 
We modeled a 432-atom supercell by repeating a 16-atom PdMn cell three times along each Cartesian axes and introduced one extra Pd atom instead of Mn in the middle of the supercell at the (0.5; 0.5; 0.5) site. Thus, the resulting composition of the defective structure is Pd$_{50.23}$Ni$_{49.77}$.
For~these two systems, we determined the lattice constants, total and site-resolved magnetic moments, and electronic densities of states~(DOS). 
The~energy cut-off for the  plane wave basis set was set to~$460$~eV. 
The~ionic relaxation was stopped when the change of the total energy became less  than~$10^{-6}$\,eV. 
The~conversion criterion for the electronic degrees of freedom  provided an accuracy of~$10^{-8}$~eV. 
The~Brillouin zone was integrated with the first-order Methfessel-Paxton~\cite{methfessel1989high} method  with a smearing parameter~$\sigma =0.1$\,eV using a Monkhorst-Pack $4\times4\times4$ $k$-point~grid.
To~show the effect of the increasing Pd excess on the electronic structure, we investigated the DOS of Pd$_{50.23}$Mn$_{49.77}$ (432~atoms) and Pd$_{56.25}$Mn$_{43.75}$ (16~atoms) and compare them with NiMn (Appendix~\ref{Appendix_DFT}).
Calculations are performed for the AF configuration depicted in  Fig.~\ref{fig_cell}, which is energetically favorable for PdMn~\cite{Yamada-2006} and NiMn alloys~\cite{Sakuma-1998, Entel-2011_MSF,Sokolovskiy-2019}.

\section{Experimental Results}
\subsection{Structure}
\label{Structure}

\begin{figure}[htbp]
\vspace*{5mm}
\includegraphics[width=0.45\textwidth]{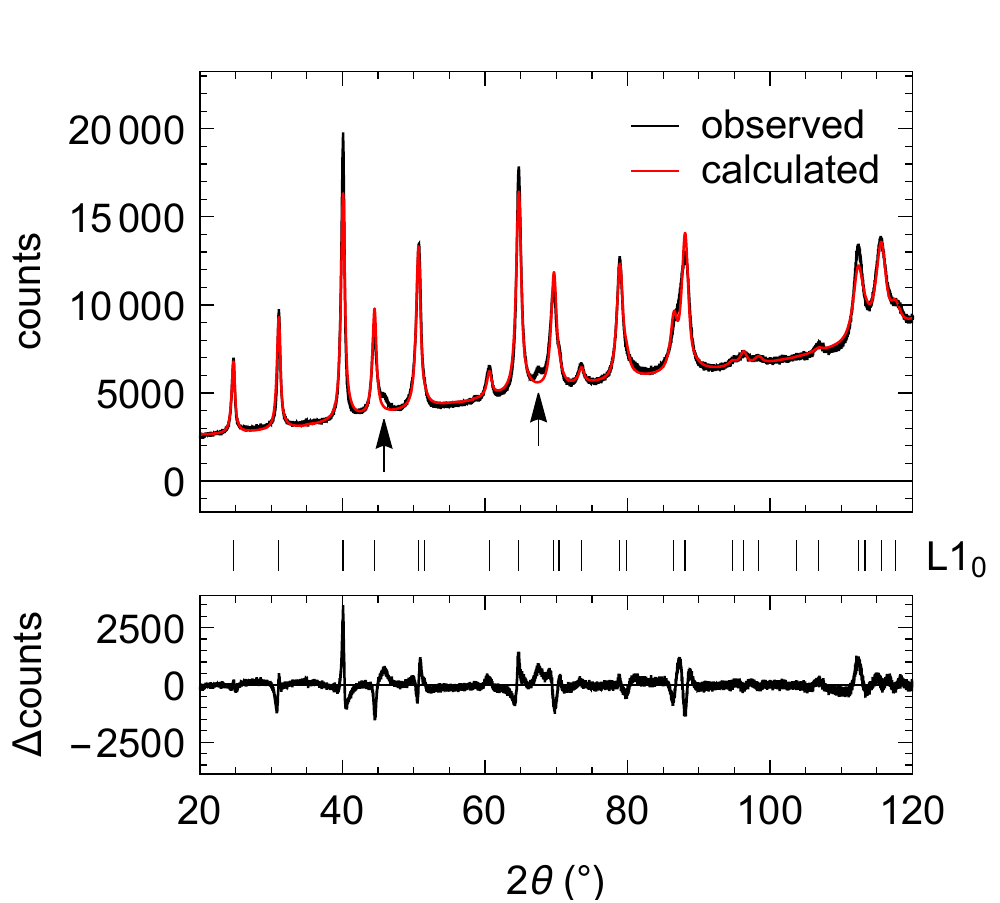}
\caption{Refinded room-temperature XRD-measurements of the bulk PdMn sample. XRD-measurements were done directly after quenching the sample from \SI{1073}{\kelvin} in room-temperature water. The measurement shows mainly the normal L1$_0$-structure of PdMn. Two additional peaks (at \ang{46} and \ang{67.5}) are visible and marked with arrows.}
\label{XRD}
\end{figure}

We show the refined XRD-data of the initial state of the PdMn sample in Fig. \ref{XRD}. The refinement was carried out with JANA2006 software \cite{petvrivcek2014crystallographic} using the P4/mmm space group. The lattice parameters were determined as $a=2.88 \pm 0.01 \text{\AA}$ and $c=3.60\pm 0.01 \text{\AA}$, which are in agreement with reference \cite{Pal2}. Two additional XRD-peaks marked with arrows are present at \ang{46} and \ang{67.5}. These could be related to the presence of Pd$_2$Mn (orthorhombic) and Pd$_3$Mn (tetragonal) \cite{miia}. The presence of these peaks are due to the unpolished edges of the sample. This is confirmed by our EDX-measurements made on polished disks cut from the original ingot showing no inhomogeneities except at the edges. Also our magnetization measurements on the cuboids, which were cut from areas away from the edges, show no features related to Pd-rich Pd-Mn \cite{grube1936magnetische}.

\subsection{Magnetization}

\subsubsection{Vertical shift in M(B)}
\label{section magnetization}
\begin{figure*}[htbp]
\vspace*{5mm}
\includegraphics[width=0.7\textwidth]{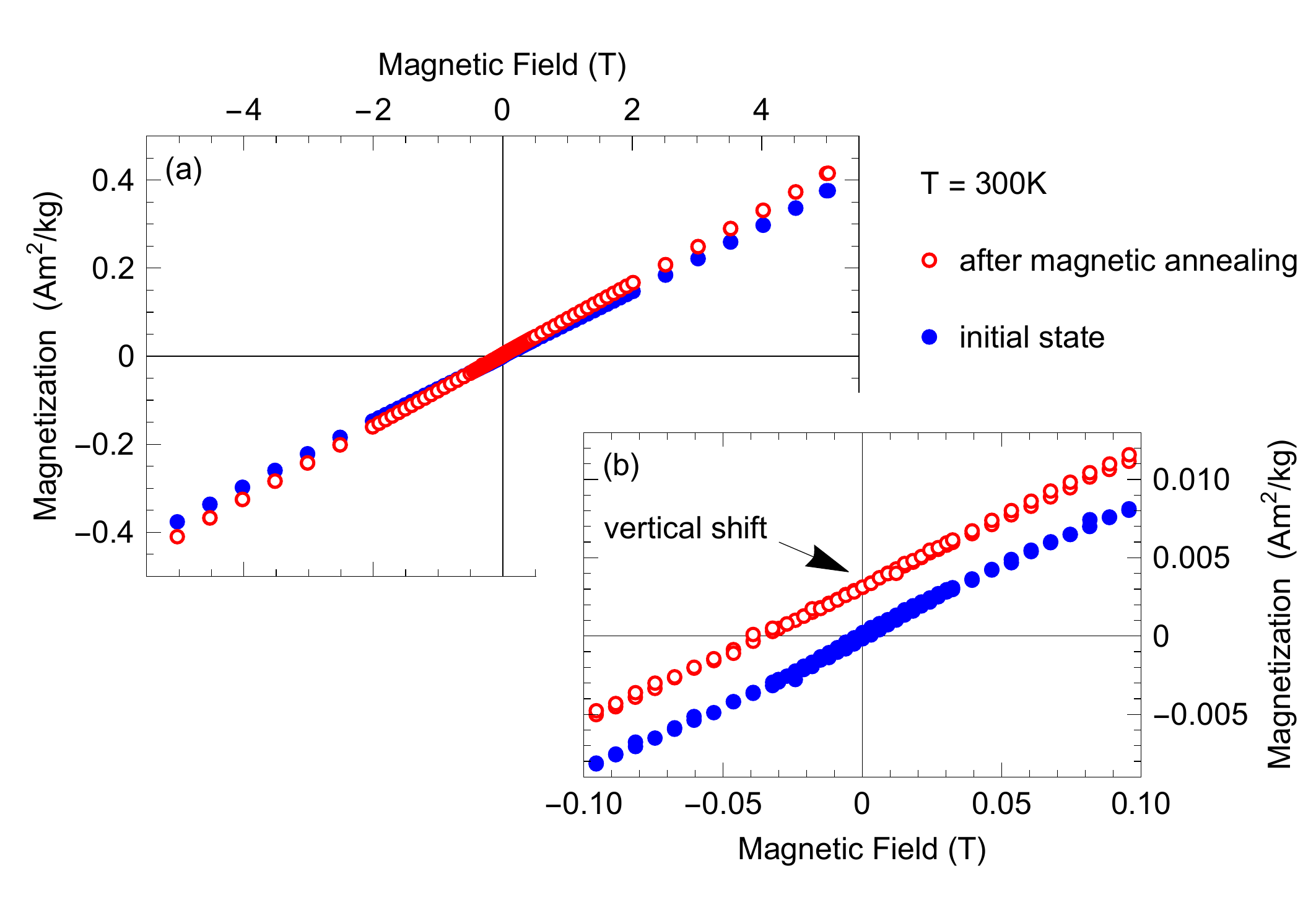}
\caption{(a) Field-dependent magnetization measurements of Pd$_{52.9}$Mn$_{47.1}$ measured at \SI{300}{\kelvin} before and after magnetic annealing for \SI{6}{\hour} at \SI{650}{\kelvin} and \SI{5}{\tesla}. After annealing the $M(B)$-curve is shifted upwards along the magnetization axis and the linear background is increased. (b) the region around zero field in more detail.}
\label{Hyst}
\end{figure*}

We show in Fig. \ref{Hyst} $M(B)$ for Pd$_{52.9}$Mn$_{47.1}$ before and after annealing for \SI{6}{\hour} at \SI{650}{\kelvin} in \SI{5}{\tesla}. The full measurement range is shown in Fig. \ref{Hyst} (a). Both $M(B)$-curves are essentially linear, which is the typical behaviour of an antiferromagnet. The magnetic susceptibility increases after magnetic annealing. 

Figure \ref{Hyst} (b) shows the region around zero field in more detail. Here, it becomes evident that the M(B)-curve is shifted vertically after magnetic annealing. In this case, the shift is ${M_\text{Shift}=(3.1\pm0.4)\times 10^{-3}\text{Am}^2\text{/kg}}$. The shift stems from pinned magnetic moments in the sample. A small FM component is also visible. It can be identified in Fig. \ref{Hyst} (b) as a S-shaped deviation from the linear dependence of the magnetization on the magnetic field. 

\subsubsection{Dependence of the vertical shift on the annealing field}
\label{field dependence section}
\begin{figure}[htbp]
\vspace*{5mm}
\includegraphics[width=0.45\textwidth]{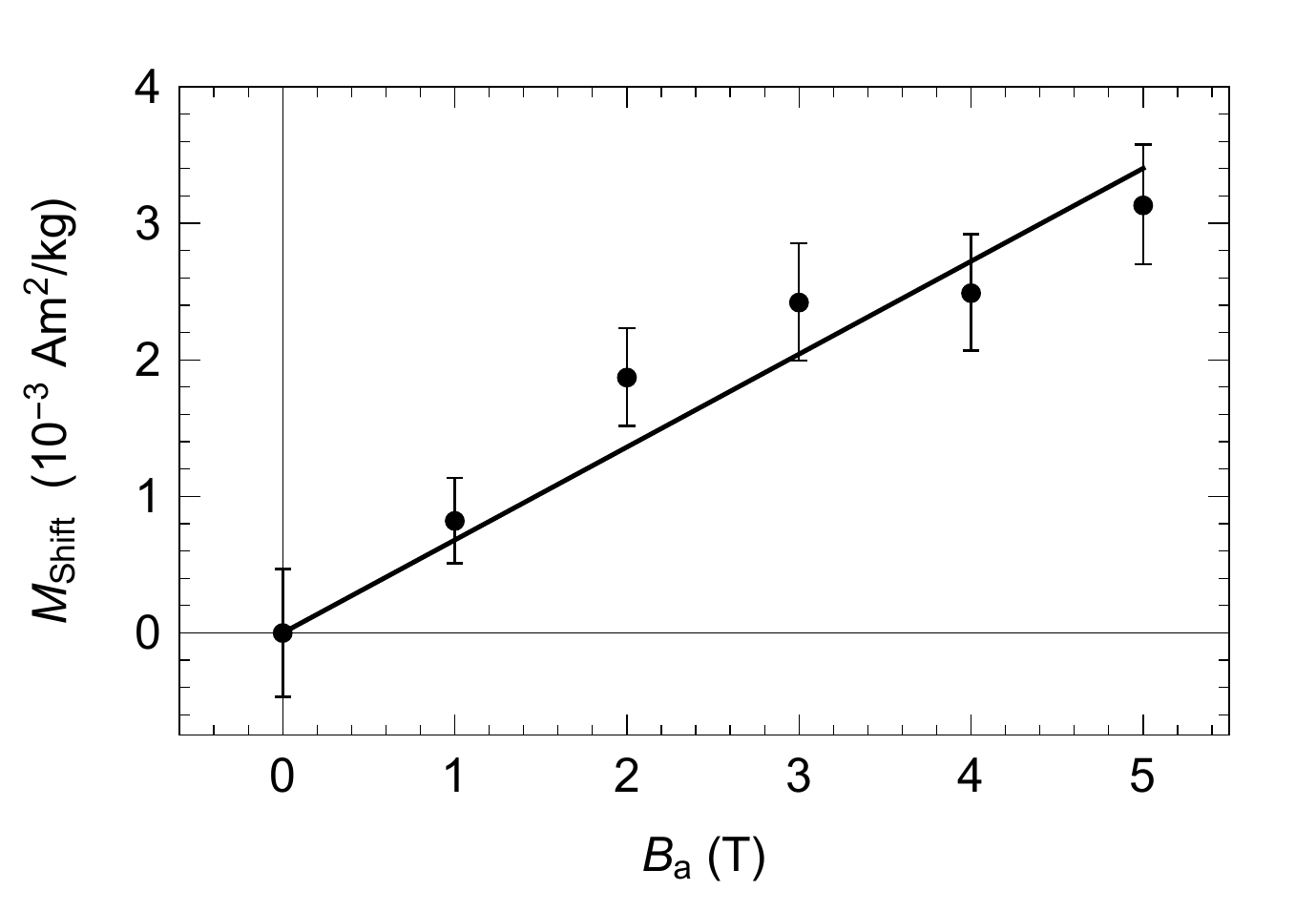}
\caption{$B_\text{a}$-dependent $M_\text{Shift}$ of Pd$_{52.9}$Mn$_{47.1}$. ${t_\text{a}=\SI{6}{\hour}}$, and the ${T_\text{a}=\SI{650}{\kelvin}}$. $B_\text{a}$ is varied between \SI{0}{\tesla} and \SI{5}{\tesla}. Each data-point represents a new annealing with a fresh sample. The vertical shifts are determined from $M(B)$ curves at \SI{300}{\kelvin}. The solid line is a linear fit.}
\label{field dependence}
\end{figure}

Figure. \ref{field dependence} shows the dependence of the size of the vertical shift on $B_\text{a}$. For each data-point, a fresh initial state sample was annealed at \SI{650}{\kelvin} for \SI{6}{\hour}. For each sample, a different $B_\text{a}$ between \SI{0}{\tesla} and \SI{5}{\tesla} was applied. The resulting vertical shift increases linearly with increasing $B_\text{a}$ with a slope of ${M_\text{Shift}=(6.8\pm1.1)\times 10^{-4}\text{Am}^2\text{/kgT}}$ . 

\subsubsection{Dependence of the vertical shift on the annealing time}
\label{time dependence section}
\begin{figure}[htbp]
\vspace*{5mm}
\includegraphics[width=0.45\textwidth]{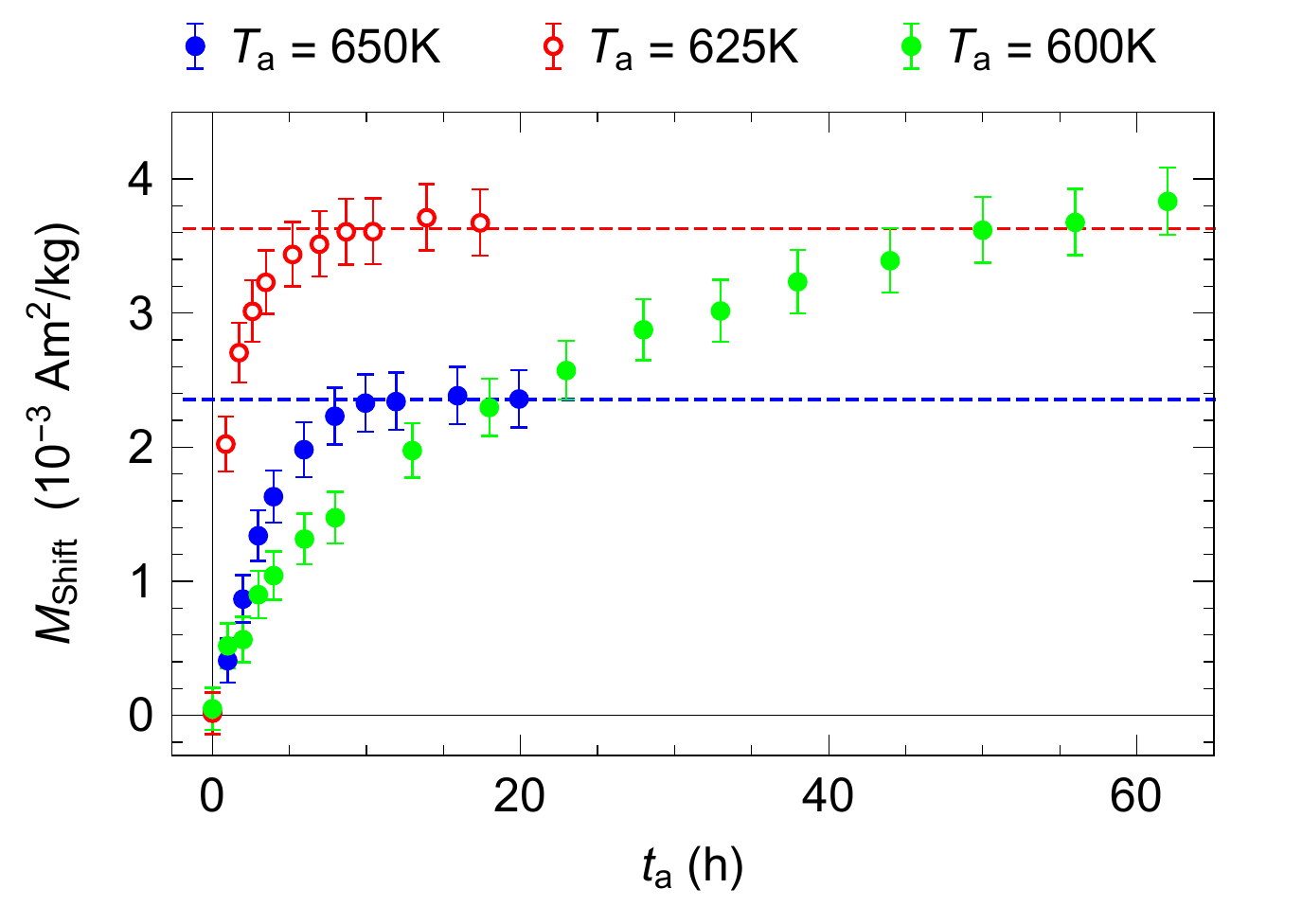}
\caption{$t_\text{a}$-dependent $M_\text{Shift}$ of Pd$_{52.9}$Mn$_{47.1}$. $B_\text{a}$ is \SI{5}{\tesla}. The time dependence was measured for values of $T_\text{a}$ of \SI{650}{\kelvin}, \SI{625}{\kelvin}, and \SI{600}{\kelvin}. For each $T_\text{a}$ the same sample was used. After measuring the full time dependence for a certain $T_\text{a}$, the sample was reset. The vertical shifts were determined from $M(B)$ curves at \SI{300}{\kelvin}. The dashed line marks the saturation value of the vertical shift for a certain $T_\text{a}$.}
\label{time dependence}
\end{figure}

Figure \ref{time dependence} shows the dependence of the size of the vertical shift on $t_\text{a}$ for values of $T_\text{a}$ of \SI{600}{\kelvin}, \SI{625}{\kelvin}, and \SI{650}{\kelvin}. $B_\text{a}$ is \SI{5}{\tesla}. The vertical shifts are determined from $M(B)$ curves obtained at \SI{300}{\kelvin}. The $t_\text{a}$-dependence is measured consecutively for each $T_\text{a}$, and for all measurements the same sample is used.

The vertical shift for all three $T_\text{a}$ increases with increasing $t_\text{a}$, while showing a tendency towards saturation. Only the time dependencies for  values of $T_\text{a}$ of \SI{625}{\kelvin} and \SI{650}{\kelvin} have been measured until full saturation, while for the \SI{600}{\kelvin} measurement $M_\text{Shift}$ does not saturate even after \SI{60}{\hour}. The vertical shift saturates for \SI{625}{\kelvin} at ${M_\text{Shift}=(3.6\pm0.24)\times 10^{-3}\text{Am}^2\text{/kg}}$ and for \SI{650}{\kelvin} at ${M_\text{Shift}=(2.4\pm0.2)\times 10^{-3}\text{Am}^2\text{/kg}}$. These values are marked with dashed lines in Fig. \ref{time dependence}. It is apparent that the saturated vertical shift for \SI{600}{\kelvin} will surpasses the values for \SI{625}{\kelvin} and \SI{650}{\kelvin}. 

To use the same sample for every $T_\text{a}$, it is necessary to do a sample-reset after each $t_\text{a}$-dependence. For a sample-reset, the sample is annealed without applying a magnetic field. This procedure sets the vertical shift back to zero. Since the $t_\text{a}$-dependence was measured at three different $T_\text{a}$ (\SI{650}{\kelvin}, \SI{625}{\kelvin}, and \SI{600}{\kelvin} in this chronological order), two resets were done. The first was \SI{4}{\hour} at \SI{650}{\kelvin} without a magnetic field and the second was \SI{4}{\hour} at \SI{650}{\kelvin} plus \SI{1.5}{\hour} at \SI{700}{\kelvin} without magnetic field.

\begin{figure}[htbp]
\vspace*{5mm}
\includegraphics[width=0.45\textwidth]{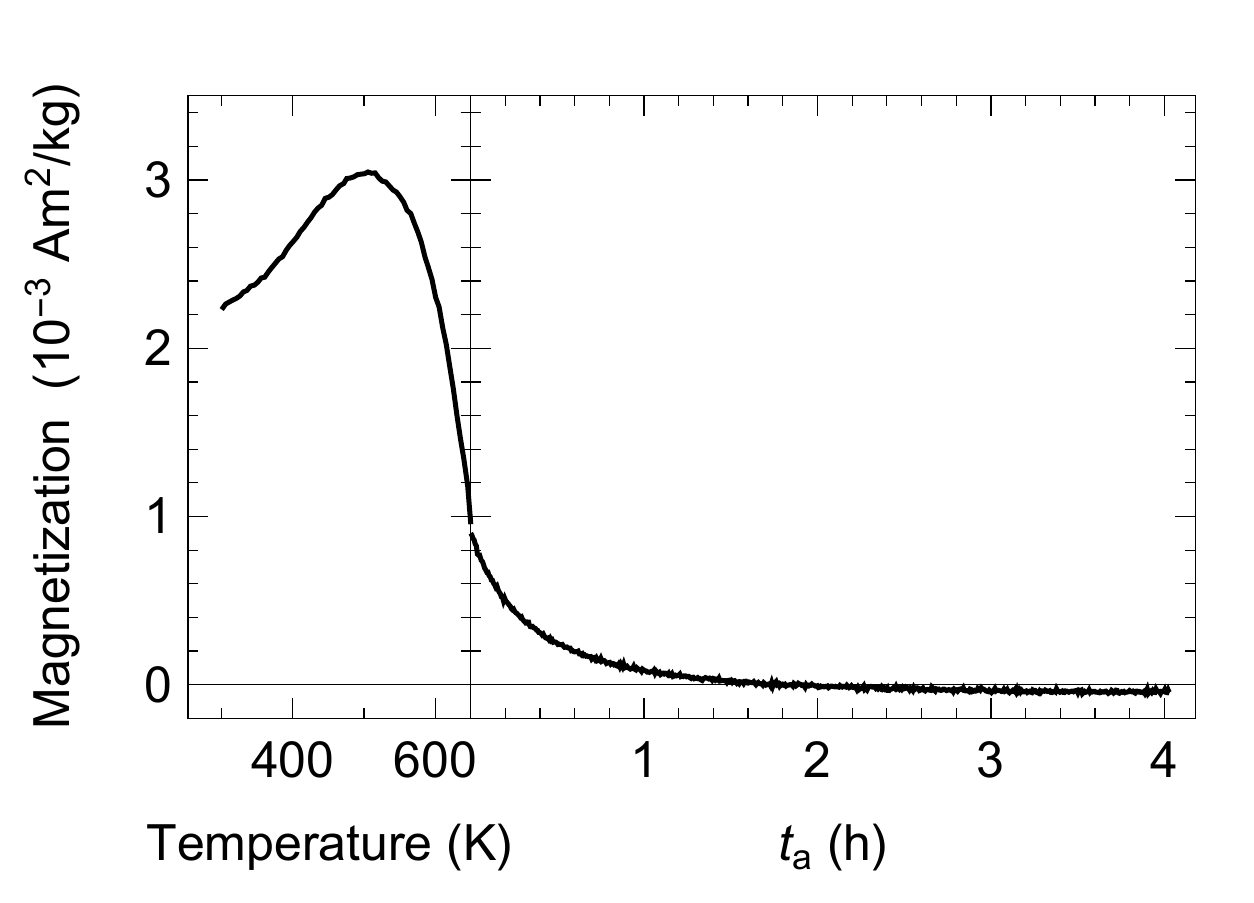}
\caption{Sample-reset done after annealing for \SI{19.9}{\hour} at \SI{650}{\kelvin} in \SI{5}{\tesla}. The plot shows a temperature dependent magnetization measurement with a speed of \SI{4}{\kelvin\per\minute} from \SI{300}{\kelvin} to \SI{650}{\kelvin} and a subsequent time dependent magnetization measurement at \SI{650}{\kelvin} for \SI{4}{\hour} all without applying a magnetic field.}
\label{reset}
\end{figure}

Such a sample-reset is shown in Fig. \ref{reset} after the first annealing procedure, which was \SI{19.9}{\hour} at \SI{650}{\kelvin} in \SI{5}{\tesla}. The measurement is divided in a temperature dependence from \SI{300}{\kelvin} to \SI{650}{\kelvin} with a rate of \SI{4}{\kelvin\per\minute} and a subsequent \SI{4}{\hour} time dependence measured with a fixed temperature of \SI{650}{\kelvin}. No magnetic field was applied during this measurement.

The measurement starts at room temperature. Since no magnetic field is applied, the measured value corresponds to the last measured vertical shift (blue in Fig. \ref{time dependence}). With increasing temperature the measured magnetization increases until reaching a maximum at around \SI{500}{\kelvin}. Afterwards, the magnetization decreases with increasing slope until \SI{650}{\kelvin}. While staying at \SI{650}{\kelvin} an exponential decay is visible until it saturates just below \SI{0}{\ampere\meter^2\per\kilo\gram} after \SI{4}{\hour}. The small negative magnetization is a result from residual currents in the superconducting magnet. We want to stress that the temperature dependence measured above \SI{500}{\kelvin} represents an non-equilibrium state so the measured value is not constant in time. Staying at \SI{550}{\kelvin} instead of \SI{650}{\kelvin} would therefore still result in an exponential decay of the magnetization. Only the time necessary to reach equilibrium would increase. After the reset, no vertical shift can be measured in the $M(B)$-curve anymore.

\subsubsection{Dependence of the vertical shift on the annealing temperature}
\begin{figure}[htbp]
\vspace*{5mm}
\includegraphics[width=0.45\textwidth]{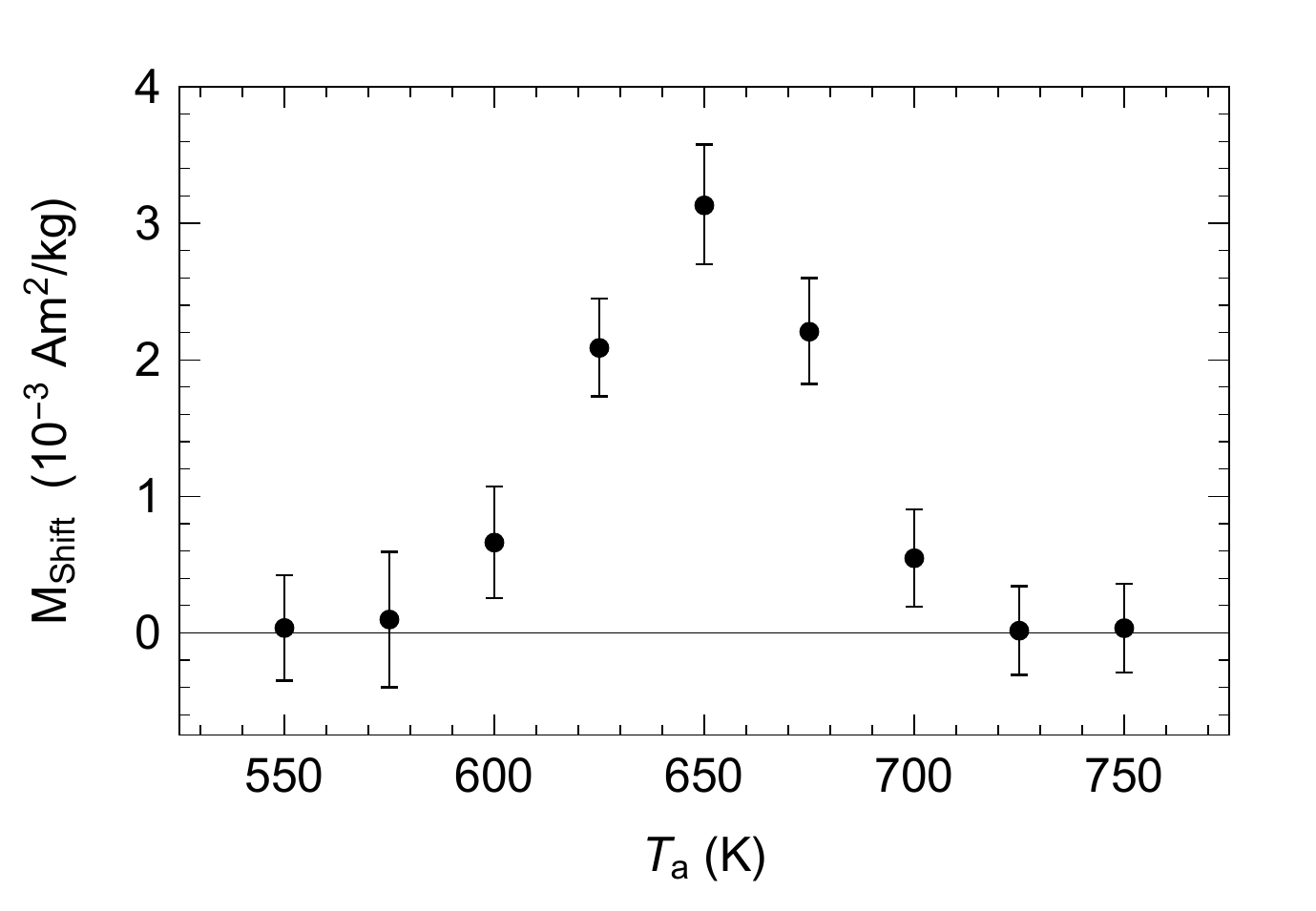}
\caption{$T_\text{a}$ dependent $M_\text{Shift}$ of Pd$_{52.9}$Mn$_{47.1}$. ${t_\text{a}=\SI{6}{\hour}}$, and $B_\text{a}$ is \SI{5}{\tesla}. $T_\text{a}$ is varied between \SI{550}{\kelvin} and \SI{750}{\kelvin}. Each data-point represents a new annealing with a fresh sample. The vertical shifts are determined from $M(B)$ curves at \SI{300}{\kelvin}}
\label{temperature dependence}
\end{figure}

We show in Fig. \ref{temperature dependence} the dependence of the size of the vertical shift on values of $T_\text{a}$ between \SI{550}{\kelvin} and \SI{750}{\kelvin}. For each data-point a fresh initial state sample was annealed in \SI{5}{\tesla} for \SI{6}{\hour}. A vertical shift begins to appear for $T_\text{a}$ above \SI{575}{\kelvin} and increases until it reaches a maximum at \SI{650}{\kelvin}. The time dependence in Fig. \ref{time dependence} has shown that this increase of $M_\text{shift}$ with increasing temperature is due to an increase in diffusion kinetics. This means $M_\text{shift}$ below \SI{650}{\kelvin} is only smaller, because it has not yet reached its equilibrium value after \SI{6}{\hour} of annealing. Above \SI{650}{\kelvin}, on the other hand, $M_\text{shift}$ decreases until it vanishes for \SI{725}{\kelvin} and above. This is because an increase in thermal energy counters the imbalance in the occupation of the two AF Mn-sublattices by the Pd-excess atoms and will therefore reduce the equilibrium value of $M_\text{shift}$.
\label{temperature dependence section}
\section{Modeling}
In this section, we calculate the dependence of the pinned magnetization on $B_\text{a}$, $t_\text{a}$, and $T_\text{a}$ based on the magnetic-field-biased diffusion model developed in references \cite{Pal,NiMnpaper} and compare it to the experimental results. For this purpose, we use the results of the \textit{ab initio} calculations. Parameters describing the diffusion kinetics were determined experimentally and are presented in App. \ref{diffusion kinetics section}.

\subsection{Magnetic properties determined from DFT}
\afterpage{\begin{figure*}[htbp]
    \centering
    \includegraphics[width=1\textwidth]{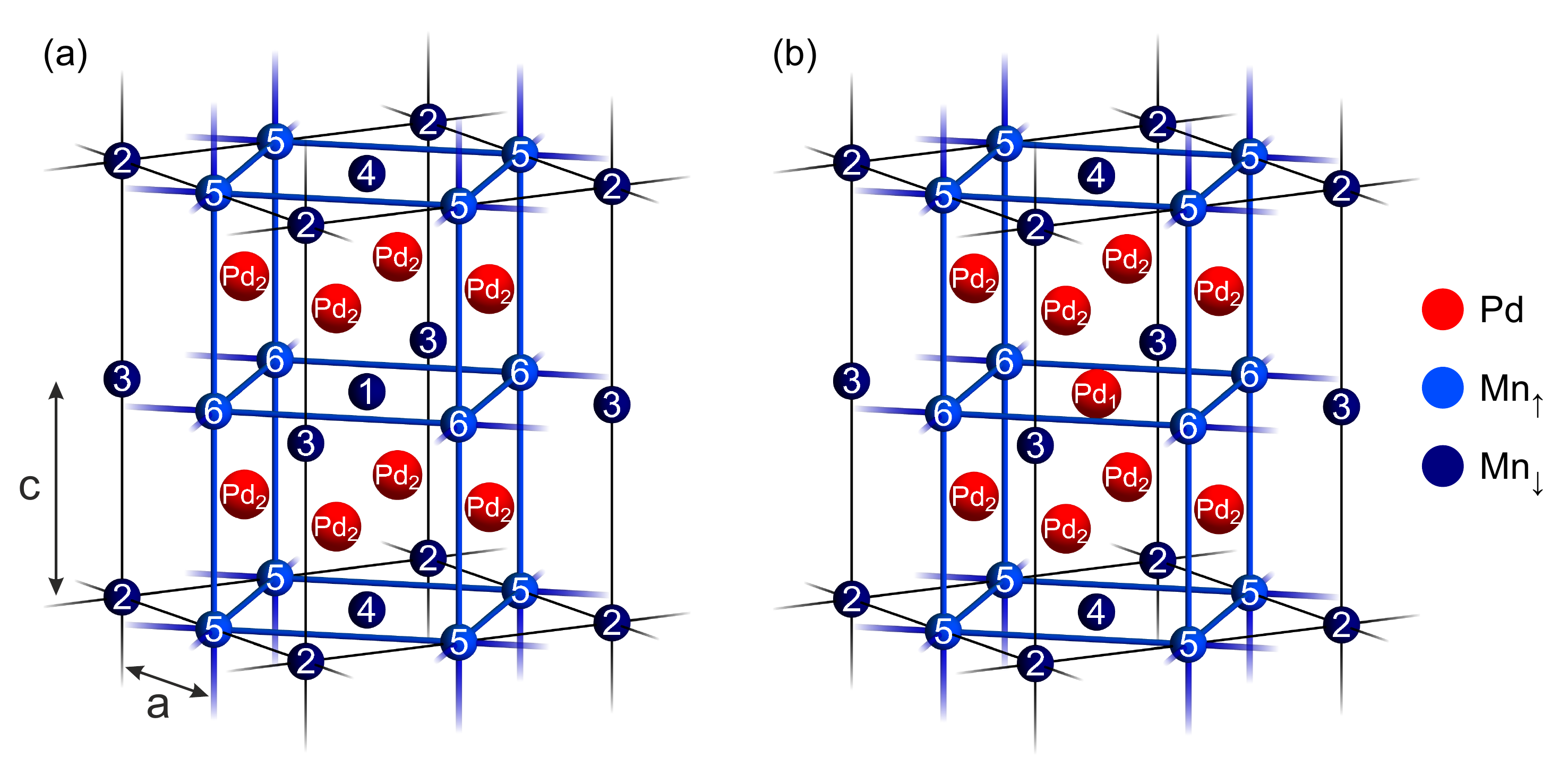}
    \caption{(a)~Equiatomic PdMn and (b)~the structure where the Mn atom in the center is replaced by a Pd atom~(Pd$_1$). This~leads to the emergence of a non-zero net magnetic moment. The two AF sublattices of the Mn atoms are indicated by two different shades of blue.
    Supercells for the first-principles calculations are formed by a $3 \times 3 \times 3$ elongation of the 16-atom cell.
    }
    \label{fig_cell}
\end{figure*}
\begin{table*}[!t]
    \caption{
    Site-resolved and total magnetic moments~$\mu$ of ideal equiatomic PdMn and the PdMn supercell with one Pd~excess~atom. 
    The~numbering of atoms corresponds to~Fig.~\ref{fig_cell}.
    }
    \label{table_magmom}
    \centering
    \begin{ruledtabular}
    \begin{tabular}{lccccccccc}
    \multirow{2}{*}{} & \multicolumn{8}{c}{Site-resolved $\mu_i$ ($\mu_\text{B}$/atom)} &   \multirow{2}{*}{Total $\mu_\text{Total}$ ($\mu_\text{B}$)} \\ \cline{2-9}
    \multicolumn{1}{c}{} & Mn$_1$ & Mn$_2$ & Mn$_3$ & Mn$_4$ & Mn$_5$ & Mn$_6$ & Pd$_1$ & Pd$_2$ &    \\ \hline
    w/o excess Pd  & -3.798 & -3.798 & -3.798 & -3.798 & 3.798 & 3.798 & $-$ & 0.000 & 0.000   \\
    with excess Pd   & $-$ & -3.777 & -3.814 & -3.824 & 3.799 & 3.858 & 0.347 & 0.030 & 4.977   \\
    \end{tabular}
    \end{ruledtabular}
\end{table*}}

We performed full geometric optimization of the PdMn and Pd-excess compositions. The~calculated lattice parameters of $a=\SI{2.82}{\angstrom}$ and $c=\SI{3.69}{\angstrom}$ agree with the experimental values in this work and Refs.~\cite{Pal2, wang2012structural} and with \textit{ab~initio} calculations~\cite{Yamada-2006}. The obtained lattice constants are nearly the same for both ideal L1$_0$-PdMn and for the structure with Pd-excess.
However, the $c/a$ ratios of these structures differ slightly, i.e., the lattice with Pd-excess is compressed in $x$ and $y$ and stretched in $c$ compared to the equiatomic structure.
This is associated with a change in the interatomic distances around the defect (Pd$_1$ atom in Fig.~\ref{fig_cell}(b)).
The nearest-neighbor~(NN) distance of $\SI{2.82}{\angstrom}$ between Mn$_1$-Mn$_6$ in PdMn decreases to $\SI{2.77}{\angstrom}$ for NN Pd$_1$-Mn$_6$ and the distance between the Mn- and Pd-planes (along $c$ axis) increases. The latter is related to the larger atomic size of the $4d$ Pd atom compared to the $3d$ Mn atoms in the plane.
This is in contrast to the NiMn system, where introducing Ni excess atoms results in a decrease of $c$, due to the, compared with Mn, smaller Ni radius.

The excess Pd atom on the Mn-position also changes the magnetic environment. Calculated site-resolved and total magnetic moments of the equiatomic and Pd-excess PdMn systems are summarised in Table~\ref{table_magmom}. In~the equiatomic composition, the Pd atoms carry no magnetic moments, while the Mn atoms each have a moment of $\pm 3.798$~$\mu_{\rm B}$. The collinear AF configuration leads to a total magnetic moment of zero.
For the Pd-excess PdMn, the Pd-excess atom in the Mn plane (Pd$_1$ in Fig.~\ref{fig_cell}(b)) results in magnetic inhomogeneities in the vicinity of this defect. NN~Mn$_6$ atoms surrounding the defect acquire the largest moments of the system of~$3.858$~$\mu_{\rm B}$.
The moments of next-NN Mn$_4$ and next-next-NN Mn$_3$ also increase in magnitude, while Mn$_5$ at the distance of $\SI{\approx 4.65}{\angstrom}$ has the same moment as in ideal PdMn.
Mn$_2$, which has the largest distance from the defect atom, has the, in absolute values, lowest magnetic moment, which is also slightly smaller than in the equiatomic composition.
This indicates that the bulk properties are not fully reached yet, and even a 432-atom cell might not be fully sufficient to harbor the entire polarization cloud.
The excess Pd atom Pd$_1$ acquires a moment of $0.347$~$\mu_{\rm B}$, which is induced by the surrounding Mn atoms and parallel to the NN Mn$_6$. The regular Pd$_2$ atoms also acquire a moment of~$0.03$~$\mu_{\rm B}$.
A Pd-excess atom on a Mn-position leads to an uncompensated configuration and results in a total magnetic moment of~$4.977$~$\mu_{\rm B}$. Compared to a Ni-excess atom in NiMn, where the total magnetic moment is~$4.997$~$\mu_{\rm B}$ \cite{NiMnpaper}, this is slightly smaller.

\subsection{Magnetic-field-biased diffusion}
\subsubsection{Dependence of the vertical shift on the annealing field}
The magnetic-field-biased diffusion model describes the experimentally observed vertical shift in $M(B)$ of Pd-rich PdMn with the accumulated excess Pd atoms in one of the two AF Mn-sublattices \cite{Pal,NiMnpaper}. The reason for this is the presence of an energy-difference between the two possible occupations of the AF Mn-sublattices in $B_\text{a}$. The vertical shift in $M(B)$ for a single crystal under equilibrium conditions, i.e., when sufficient time has elapsed after applying $B_\text{a}$, can then be calculated from  
\begin{align}
     M_\text{shift} &=  C \; \text{tanh}\left( \frac{\vec{\mu}_- \cdot \vec{B}_\text{a} }{k_\text{B} T_\text{a}} \right)\; \vec{\mu}_\text{Total} \cdot \hat{\text{e}}. \label{mass magnetization}
\end{align}
Here C is the scaling factor
\begin{align}
    C &= \frac{1}{2} \frac{\text{c}_\text{Pd}-c_\text{Mn}}{\text{c}_\text{Pd} m_\text{Pd}+c_\text{Mn} m_\text{Mn}},
\end{align}
with the atomic masses of Pd and Mn as $m_\text{Pd}$ and $m_\text{Mn}$, and their atomic concentrations as $\text{c}_\text{Pd}$ and $c_\text{Mn}$. This scaling factor is needed to obtain $M_\text{shift}$ as the mass magnetization. We use the stoichiometry of our measured sample Pd$_{52.9}$Mn$_{47.1}$ for the calculations. $\vec{\mu}_\text{Total}$ is the magnetic moment caused by a Pd-excess atom on a Mn-position in the PdMn lattice. During diffusion mediated by mono-vacancies, which is the main diffusion mechanism at the observed temperatures, $\vec{\mu}_\text{Total}$ is strongly reduced. $\mu_-$ is this reduced moment. $\hat{\text{e}}$ is a unit vector in the measurement direction, and $k_\text{B}$ the Boltzman constant. If the magnetic moments, magnetic field and the measurement direction are parallel, the Taylor expansion of Eq. \ref{mass magnetization} becomes 
\begin{align}
     M_\text{shift} &=  C \; \mu_\text{Total} \; \frac{\mu_- B_\text{a} }{k_\text{B} T_\text{a}} \;  + \mathcal{O} \left[ \left( \frac{\mu_- B_\text{a} }{k_\text{B} T_\text{a}} \right)^3  \right]. \label{Taylor series}
\end{align}
This implies that if the thermal energy is substantially larger than the Zeeman energy, a linear dependence of $M_\text{shift}$ on $B_\text{a}$ would be expected. $M_\text{shift}$ for a polycrystal has to be determined by the integration of Eq. \ref{mass magnetization} on the surface of a unit hemisphere \cite{NiMnpaper}. One obtains,
\begin{align}
\begin{split}
M_\text{shift, poly} &= C \mu_\text{Total} \\
&\frac{\pi^2-12x^2+24 x \text{ln}\left(1 +\text{exp}\left(2x \right) \right)}{24 x^2} \\
+&\frac{12 \text{Li}_2 \left(-\text{exp}\left(2x \right)  \right)}{24 x^2} \label{poly final} 
\end{split} \\
&=\frac{1}{3} \; C \; \mu_\text{Total} \; \frac{\mu_- B_\text{a} }{k_\text{B} T_\text{a}} \;  + \mathcal{O} \left[ \left( \frac{\mu_- B_\text{a} }{k_\text{B} T_\text{a}} \right)^3  \right], \label{poly final approx} 
\end{align}
where Li$_2$ is the polylogarithm of order 2. Here, we used the abbreviation 
\begin{align}
x = \frac{\mu_- B_\text{a}}{k_\text{B} T_\text{a}}.
\end{align}

For our calculations, special care has to be taken in determining the correct values for both $\mu_\text{Total}$ and $\mu_-$. $\mu_\text{Total}$ is taken from the \textit{ab initio} calculations of the equiatomic defect-free structure shown in Fig. \ref{fig_cell} (b). The corresponding value is $\mu_\text{Total}=4.977 \mu_\text{B} \approx 5 \mu_\text{B}$ from Tab. \ref{table_magmom}. In reality, the local environment of individual Pd-excess atoms can deviate from this structure. For example, multiple Pd-excess atoms can be in close proximity to each other or even neighbours. However, for practical purposes we choose this value for our calculations. $\mu_-$, on the other hand, is strongly reduced with respect to $\mu_\text{Total}$, because during mono-vacancy diffusion a Mn atom in Fig. \ref{fig_cell} (b) next to the Pd-excess atom is replaced by a vacancy and changes the local magnetic environment. Therefore, we use our experimental result to determine $\mu_-$. For this, we use the slope of the $B_\text{a}$-dependence of $M_\text{shift}$ in Sec.~\ref{field dependence section}, ${(6.8\pm1.1)\times 10^{-4}\text{Am}^2\text{/kgT}}$, and obtain $\mu_\text{Total} \times \mu_-=(1.0 \pm 0.2) \mu_\text{B}^2$. We can divide this value by $\mu_\text{Total}$ to obtain $\mu_-=0.2 \mu_\text{B}$. These values for $\mu_\text{Total}$ and $\mu_-$ will be used in further calculations.

\begin{figure}[htbp]
\vspace*{5mm}
\includegraphics[width=0.45\textwidth]{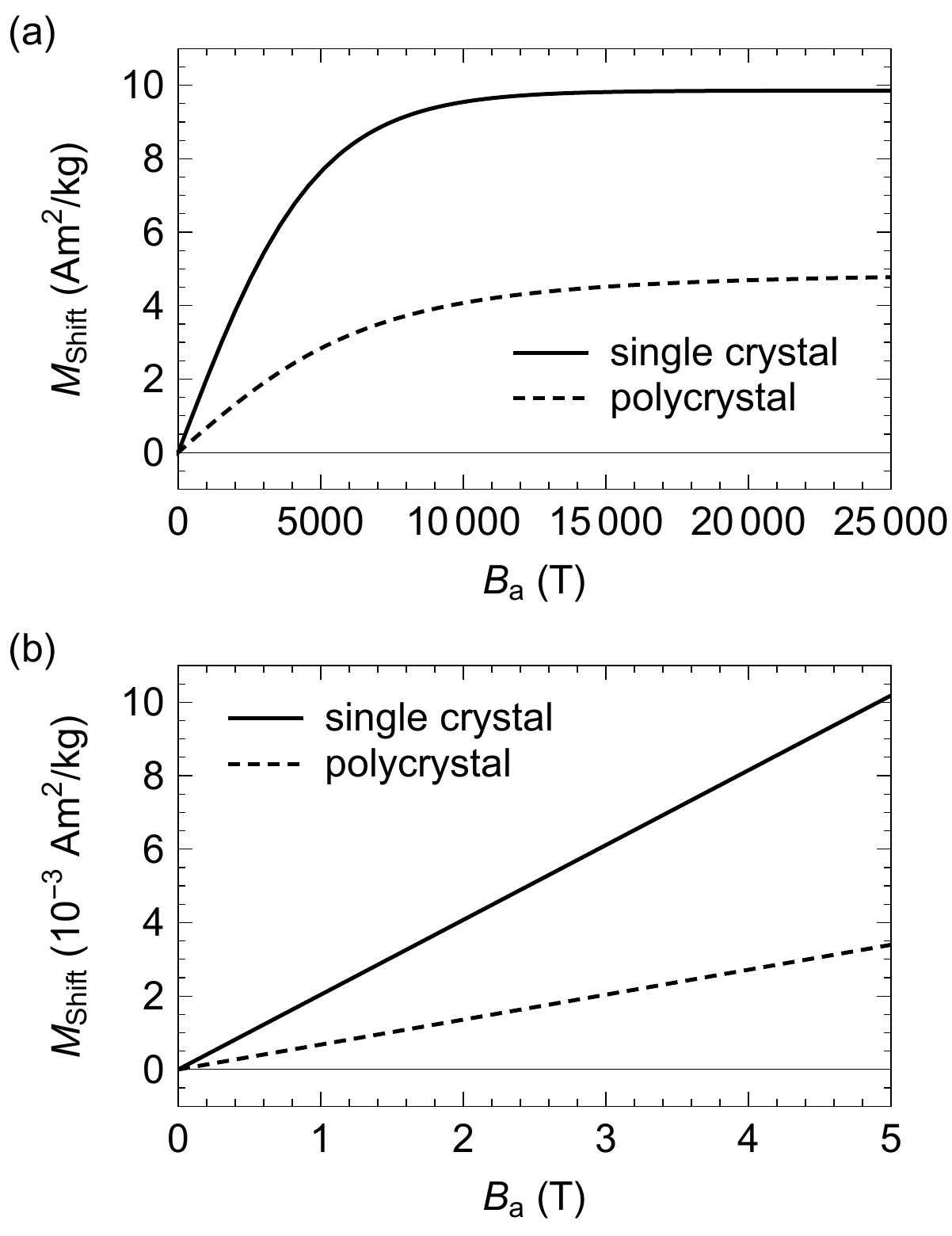}
\caption{Annealing field dependence of the vertical shift calculated for a single crystal and a polycrystal. (a) dependence up to saturation at very high hypothetical magnetic fields. (b) dependence on magnetic fields up to \SI{5}{\tesla}. For the calculations ${T_\text{a}=\SI{650}{\kelvin}}$.}
\label{field dependence calc}
\end{figure}

We show in Fig. \ref{field dependence calc} the calculated $B_\text{a}$-dependence of $M_\text{shift}$ at ${T_\text{a}=\SI{650}{\kelvin}}$. The results for both single crystals and polycrystals are plotted. In all calculations for single crystals, the magnetic field is parallel to the direction of the AF alignment. Figure \ref{field dependence calc} (a) is plotted up to very high hypothetical magnetic fields to show the saturation behaviour of $M_\text{shift}$ in Eq. \ref{mass magnetization} and Eq. \ref{poly final}. In Fig. \ref{field dependence calc} (b) we see the same calculations in fields up to \SI{5}{\tesla}. This result is the calculated counterpart of the experimental result shown in Fig. \ref{field dependence} and shows the appropriateness of the linear approximation. It is noteworthy that for small magnetic fields, $M_\text{shift}$ for a polycrystal is a third of the value for that of a single crystal (Fig. \ref{field dependence calc} (b)), while for large magnetic fields it becomes a half of the value (Fig. \ref{field dependence calc} (a)).
\subsubsection{Dependence of the vertical shift on the annealing time}
To determine the $t_\text{a}$-dependence expected from the magnetic-field-biased diffusion model, one has to consider the rate equations. The total amount of Pd-excess atoms $N=N_- + N_+$ is a fixed number. $N_-$ and $N_+$ are defined as the number of Pd-excess atoms in the sublattice with a minimized and maximized Zeeman-energy. At any point, a Pd-excess atom can change from $N_-$ to $N_+$ and vice versa. This means that the diffusion process can be treated as a reversible reaction with unequal rates. With the corresponding diffusion rates $\omega_{-}$ and $\omega_{+}$, which are given as 
\begin{align}
\omega_{\pm} &= \Gamma_0 \; \text{exp} \left( -\frac{\left( A_0 \mp 2 \; \vec{\mu}_- \cdot \vec{B}_\text{a}  \right) }{k_\text{B} T_\text{a}} \right), \label{eq omega 1}
\end{align}
one can write the rate equations as
\begin{align}
\frac{\text{d}N_-}{\text{d}t_\text{a}}&=\omega_{+}N_+-\omega_{-}N_- \\
\text{and} \qquad \qquad& \nonumber \\
\frac{\text{d}N_+}{\text{d}t_\text{a}}&=\omega_{-}N_--\omega_{+}N_+.
\end{align}
Here $\Gamma_0$ is the attempt frequency and $A_0$ the activation energy without applied field. These differential equations can be solved by separating the variables to obtain $N_- \left(t_\text{a} \right)$ and $N_+ \left(t_\text{a} \right)$. The initial conditions are ${N_- \left(0 \right)=N_+ \left(0 \right)= \nicefrac{N}{2}}$. The $t_\text{a}$-dependence of $M_\text{shift}$ is then described by $\Delta N \left( t_\text{a} \right)= N_- \left(t_\text{a} \right) - N_+ \left(t_\text{a} \right)$. Multiplying this with the scaling factor and the scalar product of $\vec{\mu}_\text{Total}$ and $\hat{\text{e}}$ gives the $t_\text{a}$-dependent magnetization as
\begin{align}
\begin{split}
M_\text{shift} \left( t_\text{a} \right) &=  C \; \text{tanh}\left( \frac{\vec{\mu}_- \cdot \vec{B}_\text{a} }{k_\text{B} T_\text{a}} \right) \\ &\Big(1-\text{exp}\left(-\left(\omega_-+\omega_+\right)t_\text{a}\right)\Big)\; \vec{\mu}_\text{Total} \cdot \hat{\text{e}}. \label{time dependence formula}
\end{split} 
\end{align}
Under experimental conditions, the Zeeman term in Eq. \ref{eq omega 1} is much smaller than the activation energy of diffusion. It can thus be neglected in calulation of $M_\text{shift}\left(t_\text{a}\right)$ so that,
\begin{align}
\omega_-+\omega_+ \approx 2 \; \omega_0 =2 \; \Gamma_0 \; \text{exp} \left( -\frac{ A_0  }{k_\text{B} T_\text{a}} \right). \label{eq omega 0}
\end{align}
With these assumptions, we obtain $\Gamma_0=\SI{1e14}{\per\second}$ and $A_0=\SI{2.3}{\evolt}$ (App. \ref{diffusion kinetics section}) using our experimental results.

\begin{figure}[htbp]
\vspace*{5mm}
\includegraphics[width=0.45\textwidth]{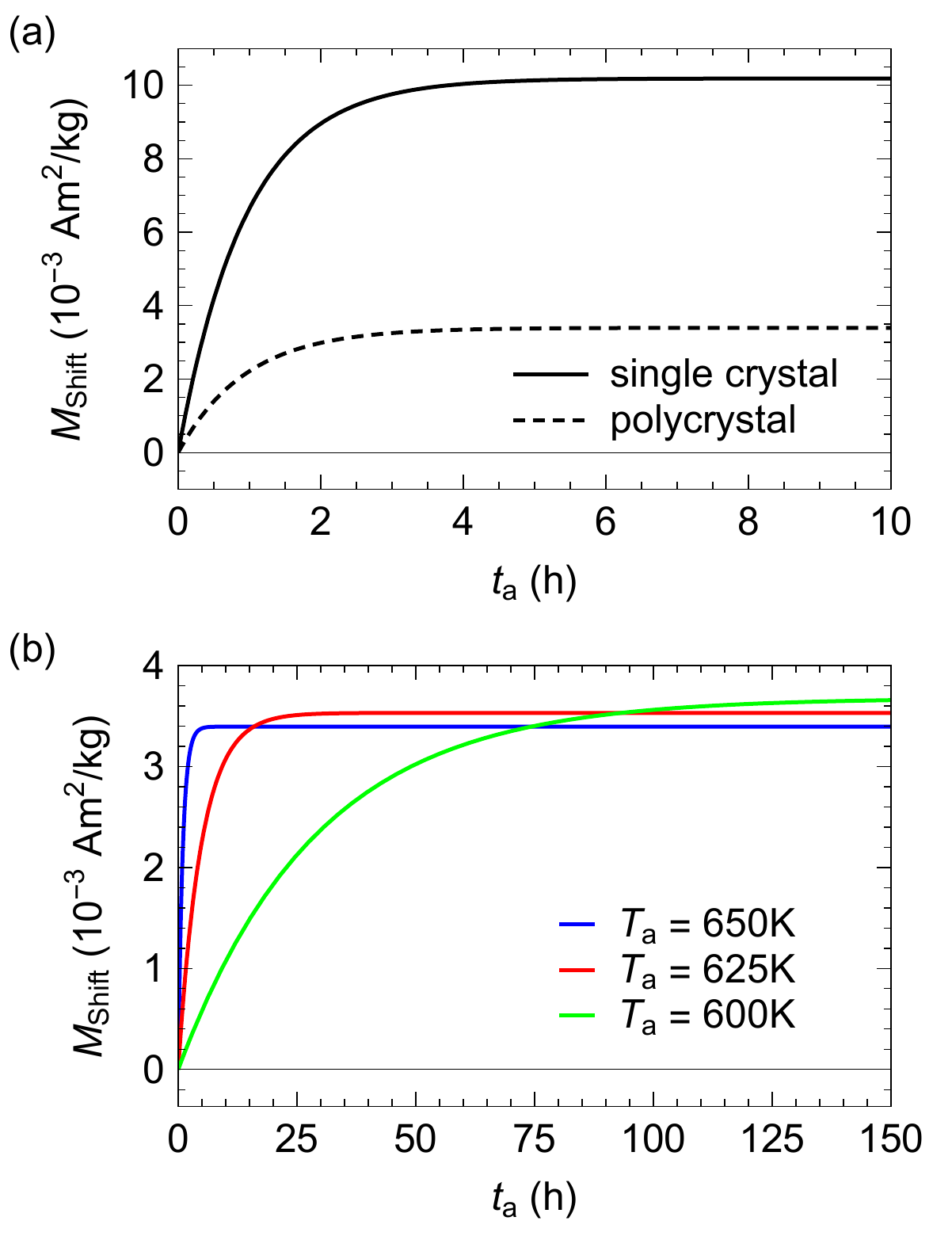}
\caption{$t_\text{a}$-dependence of the vertical shift calculated for a single crystal and a polycrystal with ${B_\text{a}=\SI{5}{\tesla}}$. (a) $t_\text{a}$-dependence at ${T_\text{a}=\SI{650}{\kelvin}}$. (b) $t_\text{a}$-dependence of a polycrystal at values of $T_\text{a}$ of \SI{600}{\kelvin}, \SI{625}{\kelvin} and \SI{650}{\kelvin}.}
\label{time dependence calc}
\end{figure}

We show in Fig. \ref{time dependence calc} the calculated $t_\text{a}$-dependencies of $M_\text{shift}$. In Fig. \ref{time dependence calc} (a), we compare the results for single crystals and for polycrystals at ${T_\text{a}=\SI{650}{\kelvin}}$ and ${B_\text{a}=\SI{5}{\tesla}}$. In Fig. \ref{time dependence calc} (b) we show the $t_\text{a}$-dependence of $M_\text{shift}$ for the three temperatures \SI{600}{\kelvin}, \SI{625}{\kelvin} and \SI{650}{\kelvin}. With higher temperature, the rate of increase of $M_\text{shift}$ increases with higher $T_\text{a}$, while the saturation value of $M_\text{shift}$ decreases. This is the calculated counterpart of the experimental result shown in Fig. \ref{time dependence}. 
\subsubsection{Dependence of the vertical shift on the  annealing temperature}
The dependence of $M_\text{shift}$ on $T_\text{a}$ can be calculated in the same way as the dependence of $M_\text{shift}$ on $B_\text{a}$ using Eq. \ref{mass magnetization} and Eq. \ref{poly final}. 

\begin{figure}[htbp]
\vspace*{5mm}
\includegraphics[width=0.45\textwidth]{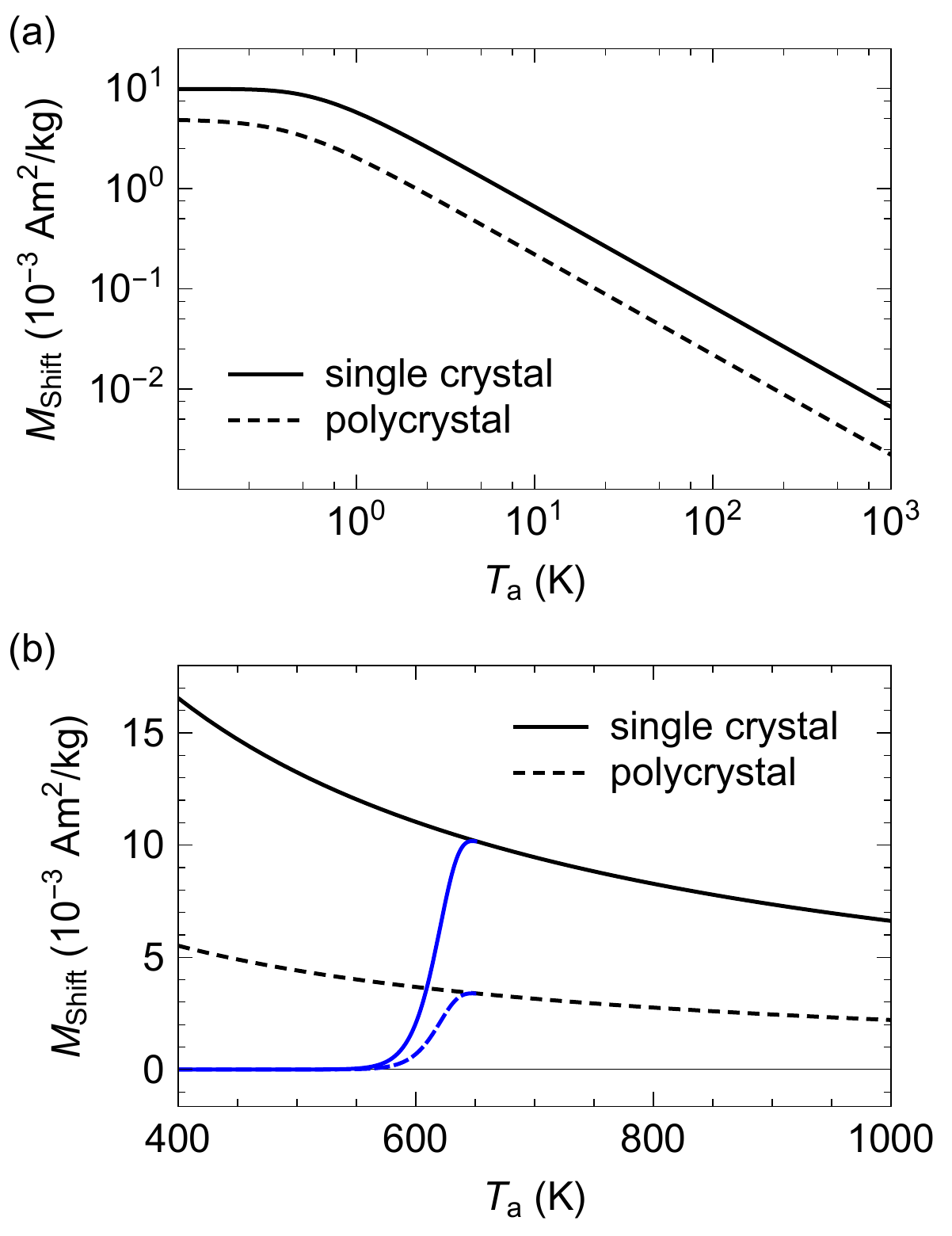}
\caption{Annealing temperature dependence of the vertical shift calculated for a single crystal and a polycrystal in ${B_\text{a}=\SI{5}{\tesla}}$. (a) log-log plot of the vertical shift in equilibirum. (b) plot of the vertical shift in equlibrium (black) and in an intermidiate state after \SI{6}{\hour} of annealing (blue).}
\label{temperature dependence calc}
\end{figure}

We show in Fig. \ref{temperature dependence calc} the calculated temperature dependence of $M_\text{shift}$ for ${B_\text{a}=\SI{5}{\tesla}}$ for a single crystal and a polycrystal.
Figure \ref{temperature dependence calc} (a) shows $M_\text{shift}$ in the temperature range ${\SI{0.1}{\kelvin} \leq T_\text{a} \leq \SI{1000}{\kelvin}}$ in a log-log plot. Here, the inverse proportionality of $M_\text{shift}$ on $T_\text{a}$ for temperatures higher than \SI{1}{\kelvin} is visible, which is a result of the dominating term in the Taylor series expansions shown in Eq. \ref{Taylor series} and Eq. \ref{poly final approx}. For lower temperatures, $M_\text{shift}$ for both single crystals and polycrystals saturates at values equal to their respective saturation values in Fig. \ref{field dependence calc} (a). This corresponds to a full occupation of Pd-excess atoms in one of the AF sublattices. This can only be achieved by annealing at very high fields or at very low temperatures. At high $T_\text{a}$, $M_\text{shift}$ for a polycrystal is a third of the value of that for a single crystal. At low $T_\text{a}$, it is a half. This behavior is the opposite of that observed for the dependence of $M_\text{shift}$ on $B_\text{a}$.

In Fig. \ref{temperature dependence calc} (a), $M_\text{shift}$ only at equilibrium conditions is plotted. In an experiment, $t_\text{a}$ is finite and thus equilibrium at lower $T_\text{a}$ with small diffusion rates cannot be achieved. This means that the dependence of $M_\text{shift}$ on $T_\text{a}$ obtained in an experiment will always deviate from the result calculated for equilibrium. To account for this, $M_\text{shift}$ has to be calculated for an intermediate state after a finite $t_\text{a}$ using Eq. \ref{time dependence formula}. Figure \ref{temperature dependence calc} (b) shows the $T_\text{a}$-dependence of $M_\text{shift}$ in a temperature range between \SI{400}{\kelvin} and \SI{1000}{\kelvin} for both the equilibrium value ($t_\text{a} \rightarrow \infty$, black) and an intermediate state ($t_\text{a} = \SI{6}{\hour}$, blue). Again, both the results for a single crystal and a polycrystal are shown. This is the calculated counterpart of the experimental result shown in Fig. \ref{temperature dependence}. The decrease of $M_\text{shift}$ below \SI{650}{\kelvin} is also visible from the experimental results (Fig. \ref{temperature dependence} and the connection to decreased diffusion rates is confirmed by our $t_\text{a}$-dependent measurements in Sec. \ref{time dependence section}. The decrease of $M_\text{shift}$ with increasing $T_\text{a}$ above \SI{650}{\kelvin}, on the other hand, is much stronger than the calculated inverse proportional dependence on $T_\text{a}$ and is also observed in the $t_\text{a}$-dependent measurements.

\section{conclusions}
Magnetic annealing of PdMn with excess Pd leads to strongly pinned uncompensated magnetic moments, which are noticeable as a vertical shift in $M(B)$. A model to explain this phenomenon is the magnetic-field-biased diffusion model, which was introduced within the context of studies on the magnetic properties of NiMn \cite{Pal,NiMnpaper}. The vertical shift in PdMn is of the same magnitude as for NiMn.

We measured the annealing time, temperature and field dependence of the vertical shift in $M(B)$ of Pd$_{52.9}$Mn$_{47.1}$ and compared the results with those of the magnetic-field-biased diffusion model. For this we used \textit{ab initio} calculations and determined the magnetic moment caused by a Pd-excess atom on a Mn position 
as $4.977$~$\mu_\text{B}$. This includes the absent moment of the replaced Mn atom, which would have pointed in the opposite direction, the induced moment of the Pd-excess atom itself, and minor contributions from the surrounding magnetic environment. 
This is close to the previously discussed case of NiMn~\cite{NiMnpaper}.
We confirm a good agreement between model and experiment. However, the model deviates from the experimental results for $T_\text{a}$ higher than \SI{650}{\kelvin}.

\section*{Acknowledgments}
We acknowledge funding by the German Research Foundation (DFG) within the Collaborative Research Center/Transregio (CRC/TRR) 270 (Project-No. 405553726, subprojects A04 and B06). The authors gratefully acknowledge the computing time provided to them at the NHR Center NHR4CES at TU~Darmstadt (project number~p0020039).
This~is funded by the Federal Ministry of Education and Research, and the state governments participating on the basis of the resolutions of the GWK for national high performance computing at universities (www.nhr-verein.de/unsere-partner).
We~thank Ulrich Nowak (University of Konstanz) and Alfred Hucht (University of Duisburg-Essen) for helpful discussions.
\appendix

\section{Effect of Pd excess on the electronic structure }
\label{Appendix_DFT}
\begin{figure}[t]  
    \centering
    \includegraphics[width=0.45\textwidth]{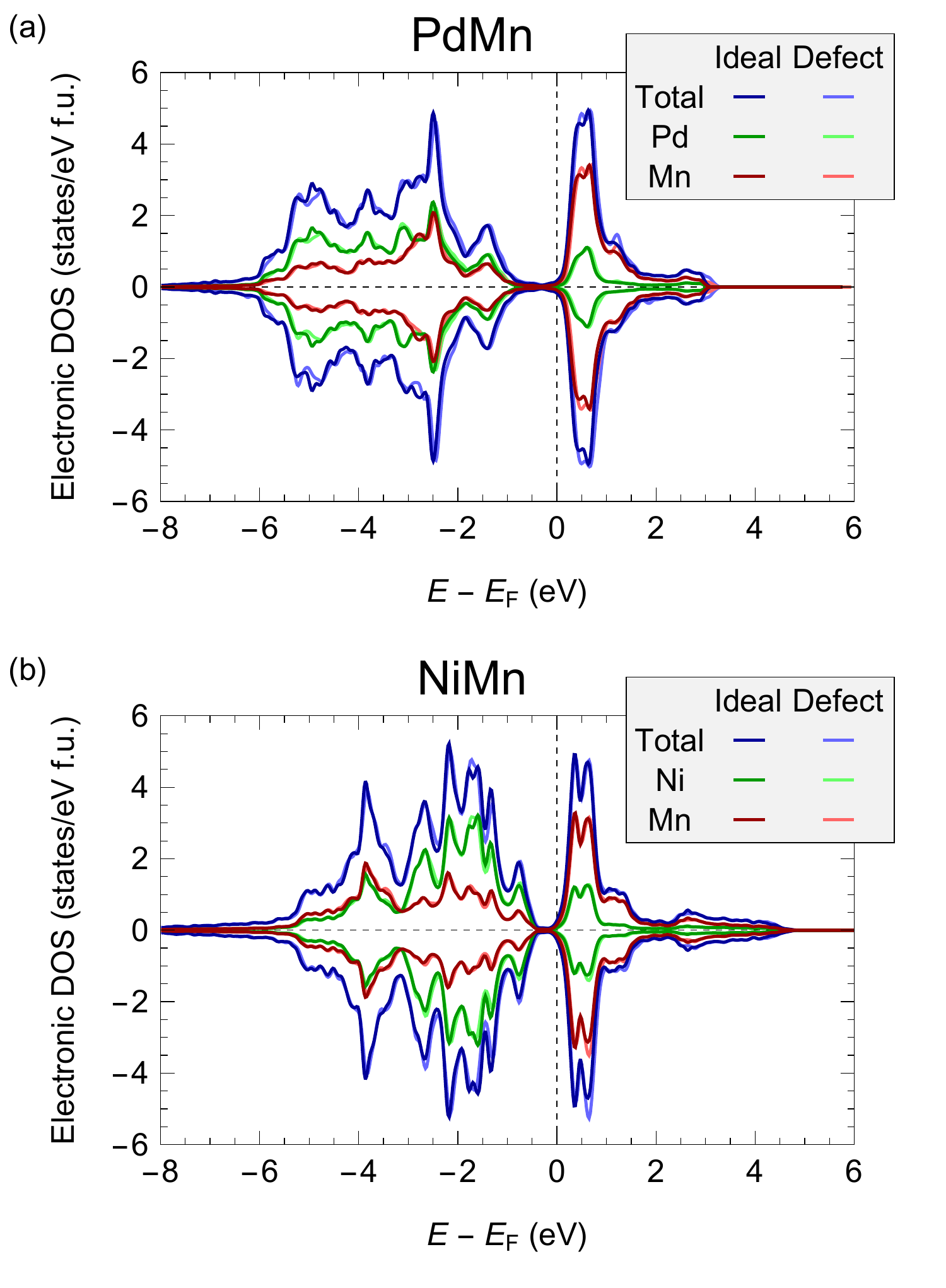}
    \caption{Total and element-resolved electronic DOS of (a)~PdMn and (b)~NiMn in comparison with corresponding defective structures. Results are obtained for 432-atom supercells and normalized to 2 atoms. Defective structures contain one extra Pd/Ni atom introduced in the middle of the supercell (see the detailed description in Sec.~II~B).
    }
    \label{Fig_DOS}
\end{figure}

To get a better understanding of the differences between PdMn and NiMn, we calculated total and partial electronic DOS of the equiatomic AF L1$_0$-PdMn and the structure with Pd excess and compare them with the analogous structures of NiMn, which have the same AF configuration.
The total DOS profile of PdMn (Fig.~\ref{Fig_DOS}(a)) agrees with the calculation in Ref.~\cite{wang2012structural}.
For both ideal and defective structures, the peak at $\approx -2.5$~eV is formed by strong Pd-Mn hybridization with equal contributions from Mn and Pd atoms.
The two lower features at $\approx -3.9$ and $-4.9$~eV have an almost twice as large contribution from Pd, which can be attributed to the larger number of valence electrons of~Pd.
The latter peak ($-4.9$~eV) shows the most noticeable difference between the equiatomic and Pd-excess structure, since it is replaced by a local minimum at $-5$~eV in the Pd-excess structure. Both Pd systems are characterised by a deep minimum in the DOS around the Fermi level~$E_{\rm F}$ forming a pseudogap.
This has been reported previously for L1$_0$-type ordered Mn$_{1-x} Z$, where $Z=$~Ni~\cite{Sakuma-1998}, Pt~\cite{Kubota-2007}, and Pd~\cite{Umetsu-2006, umetsu2006effective, Kubota-2007}.
Passing from Pd- to Ni-systems results in the narrowing of this pseudogap from 1.17~eV to~0.58~eV.


Of particular interest for us is, how further increasing the defect concentration would affect the electronic structure. 
Therefore, we additionally performed the full relaxation of a system with 6.25~at.\% Pd excess. We~found a similar trend as obtained by Umetsu~\textit{et~al.}~\cite{umetsu2006effective}, who used a tight-binding linear muffin-tin orbital~(LMTO) method with the coherent potential approximation~(CPA), which does not involve lattice relaxation.
Counterintuitively, the changes cannot simply be described by the rigid band model, where the additional $d$-electrons shift the Fermi level~$E_{\rm F}$ and increasing Pd-excess would be expected to shift the Fermi level into the larger peak above the pseudogap.
However, The calculation shows that weight from the unoccupied DOS is redistributed below the pseudogap, and $E_{\rm F}$ finally locates at the bottom edge of the~gap.
As~a consequence, the occupied $3d$ states move in total closer to the Fermi level, which is associated with an increase in band energy.
We~speculate that this suppresses clustering of the excess Pd when the overall Pd concentration is low enough to keep $E_{\rm F}$  close to the upper edge of pseudogap and, therefore, supports the field-induced diffusion of the excess-Pd atoms.
The situation is similar to NiMn, where the pseudogap is, however, two times smaller.

\begin{figure}[htbp]
    \centering
    \includegraphics[width=0.5\textwidth]{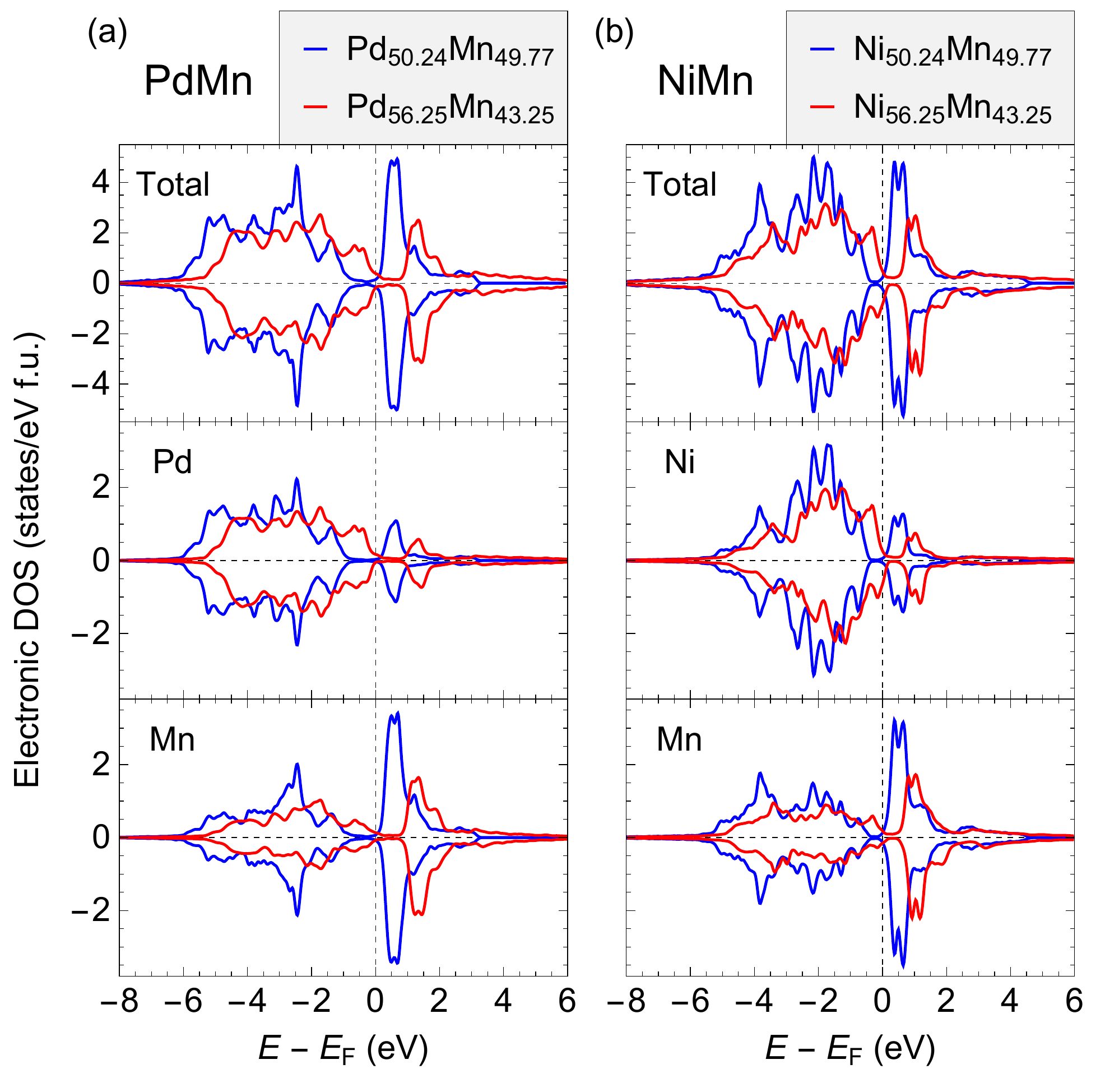}
    \caption{Total~(first row) and element-resolved (second and third rows for Pd/Ni and Mn, correspondingly) electronic DOS of Pd-Mn~(left panel) and Ni-Mn~(right panel).
    Passing from smaller (blue curves) to larger (red curves) Pd/Ni-excess content, increasing of valence electron concentration results in shifting of the whole DOS weight towards higher energies pushing the pseudo-gap above the Fermi level.
    }
    \label{Fig_DOS_16at}
\end{figure}

\section{Diffusion kinetics}
\label{diffusion kinetics section}
\begin{figure}[htbp]
\vspace*{5mm}
\includegraphics[width=0.45\textwidth]{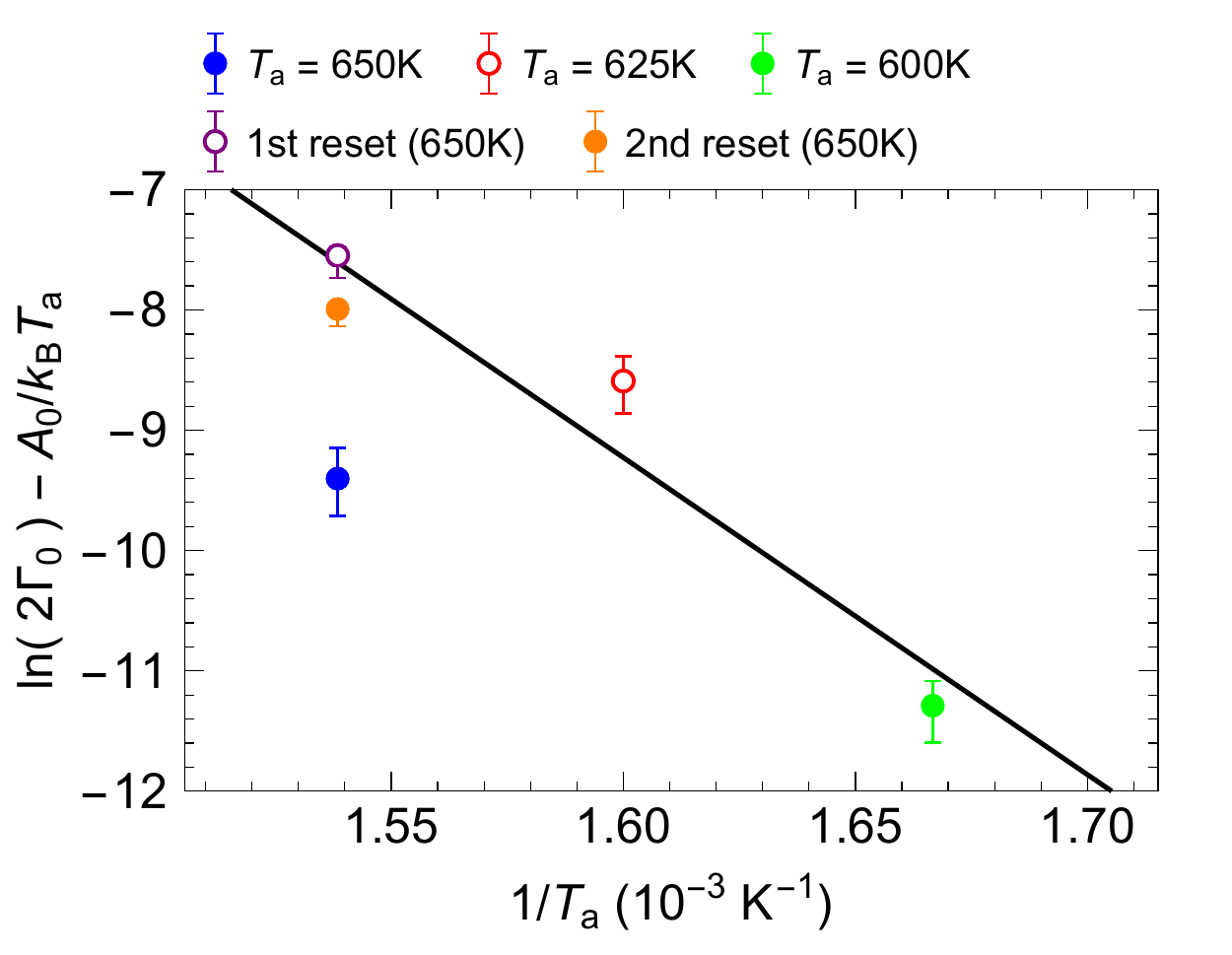}
\caption{${\text{ln}\left(\Gamma_0 \right)-A_0/k_\text{B}T_\text{a}}$ plottet against $1/T_\text{a}$. The solid line is a linear fit from which $\Gamma_0$ and $A_0$ can be determined.}
\label{attempt and activation}
\end{figure}

For modeling the $t_\text{a}$- and $T_\text{a}$-dependence of $M_\text{shift}$, the attempt frequency $\Gamma_0$ and the activation energy $A_0$ are needed. An estimation of these values can be obtained from our $t_\text{a}$-dependent measurements of $M_\text{shift}$. In total, we measured five $t_\text{a}$-dependencies, which are discussed in Sec. \ref{time dependence section}. If the Zeeman-contribution is neglected, these $t_\text{a}$-dependencies can all be described by,
\begin{align}
\begin{split}
    \Delta M_\text{shift}&\left(t_\text{a},T_\text{a}\right) = \\ &\text{exp}\left( \left[-2 \; \Gamma_0 \; \text{exp}\left(- \frac{A_0}{k_\text{B} T_\text{a}} \right) \right] t_\text{a} \right). \label{time formula}
\end{split}
\end{align}
$\Delta M_\text{shift}$ is the normalized difference in pinned magnetization between intermediate and equlilibrium states. If the $t_\text{a}$-dependence is exponential, then the natural logarithm of $\Delta M_\text{shift}$ should depend linearly on $t_\text{a}$.
\begin{figure}[htbp]
\vspace*{5mm}
\includegraphics[width=0.45\textwidth]{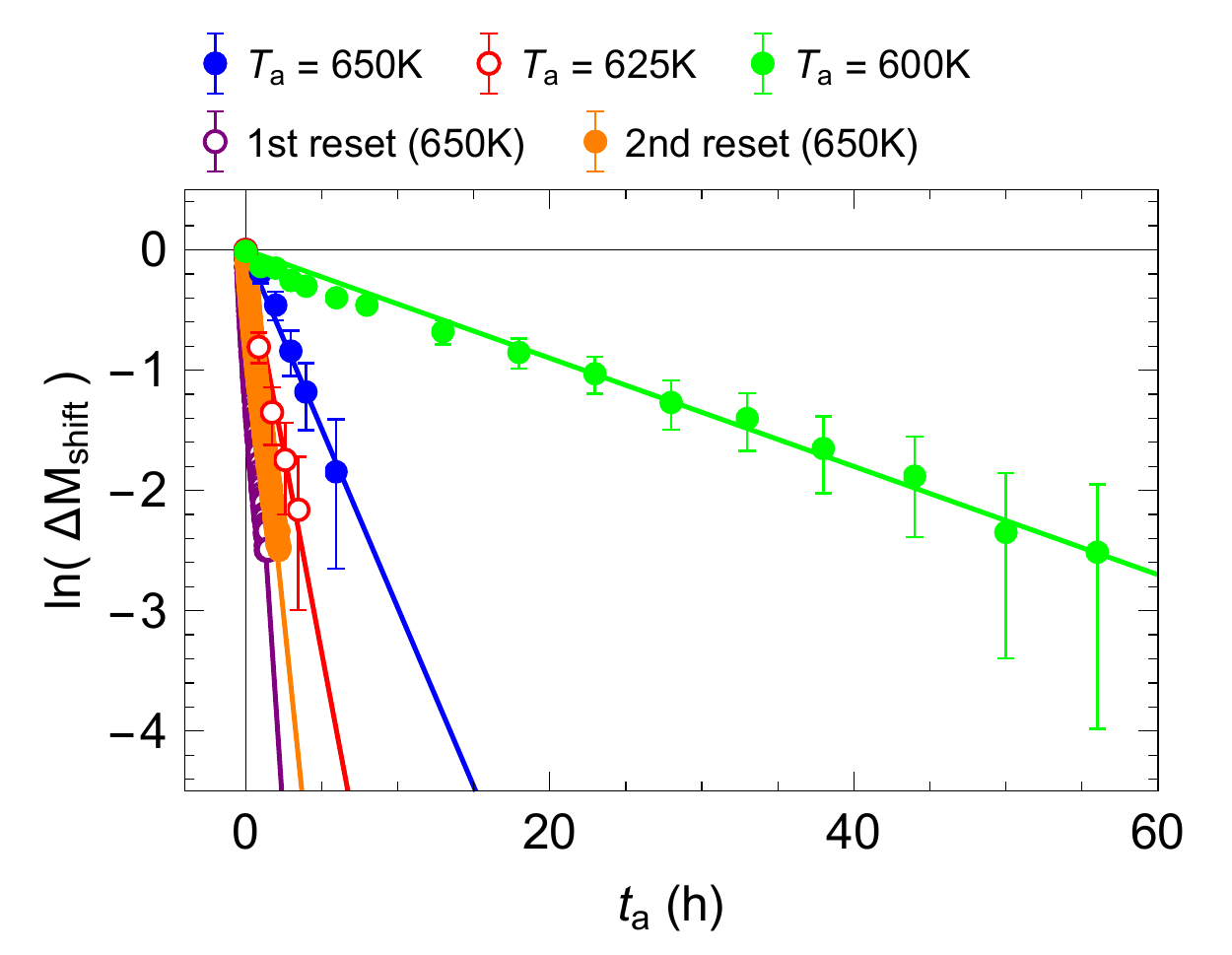}
\caption{The natural logarithm of $\Delta M_\text{shift}$ for three $T_\text{a}$ plotted against $t_\text{a}$. A linear dependence indicates an exponential $t_\text{a}$-dependence of $\Delta M_\text{shift}$. The respective linear fits are shown in the same color as the data-points.}
\label{Logplot}
\end{figure}
This dependence is shown in Fig. \ref{Logplot}. Following Eq. \ref{time formula}, the slopes of ln$(\Delta M_\text{shift})$ should become steeper with increasing $T_\text{a}$. Here, $T_\text{a}=\SI{650}{\kelvin}$ is an outlier, which does not follow this rule. This could be 
be associated with the sequential order of the chosen $T_\text{a}$. In the experiments, $T_\text{a}=\SI{650}{\kelvin}$ was measured first, followed by $T_\text{a}=\SI{625}{\kelvin}$ and $T_\text{a}=\SI{600}{\kelvin}$. This means that the reduced diffusion rate at $T_\text{a}=\SI{650}{\kelvin}$ could be caused by lattice relaxations or the release of stress inside the sample during the first annealing. We, therefore, do not take this measurement into account when determining the attempt frequency and the activation energy. The slopes of the linear fits shown in Fig. \ref{Logplot} correspond to the factor in front of $t$ in Eq. \ref{time formula}. The natural logarithm of the slopes, multiplied by -1, then gives ${\text{ln}\left(\Gamma_0 \right)-A_0/k_\text{B}T_\text{a}}$, which can be plotted against $1/T_\text{a}$. The results can be seen in Fig. \ref{attempt and activation}.
Again, the value for the first annealing at $T_\text{a}=\SI{650}{\kelvin}$ is an outlier. $\Gamma_0$ and $A_0$ can now be determined by a linear fit excluding this point. We obtain $\Gamma_0=\SI{1e14}{\per\second}$ and $A_0=\SI{2.3}{\evolt}$. No error is given since this is just an estimation. We now compare these results with those in the literature. In magnetic-field-biased diffusion of NiMn, an activation energy of $A_0=\SI{1.8}{\evolt}$ was determined \cite{Pal}, which is in the same range as our result. In reference \cite{peterson1964isotope}, the diffusion coefficient of pure palladium is found to be $D_0=\SI{0.205}{\square\centi\meter\per\second}$. This value has to be translated into an attempt frequency by calculating
\begin{align}
    \Gamma_0 = \frac{D_0}{f \; a_\text{Pd}^2}.
\end{align}
Here, $f$ is the correlation factor, which is 0.7815 for vacancy diffusion and $a_\text{Pd}$ is the lattice parameter of Pd. This results in $\Gamma_0=\SI{1.7e14}{\per\second}$, which is also of the same magnitude as our result.

\newpage
~

\end{document}